\documentclass[aps,prd,twocolumn,superscriptaddress,preprintnumbers,amsmath,amssymb,nofootinbib]{revtex4-1}

\usepackage{hyperref}
\usepackage{graphicx}
\usepackage{xcolor}
\usepackage{bm}

\usepackage[utf8]{inputenc}

\usepackage{setspace}
\usepackage{footmisc}
\allowdisplaybreaks

\begin{document}

\preprint{ULB-TH/16-11, DESY 16-127}

\title{Dark matter and observable Lepton Flavour Violation}
\author{Lucien Heurtier}
\email{lucien.heurtier@ulb.ac.be}

\affiliation{Service de Physique Th\'eorique - Universit\'e Libre de Bruxelles, Boulevard du Triomphe, CP225, 1050 Brussels, Belgium}

\affiliation{Deutsches Elektronen-Synchrotron DESY, 22607 Hamburg, Germany}

\author{Daniele Teresi}
\email{daniele.teresi@ulb.ac.be}

\affiliation{Service de Physique Th\'eorique - Universit\'e Libre de Bruxelles, Boulevard du Triomphe, CP225, 1050 Brussels, Belgium}

%\date{\today}

\begin{abstract}
Seesaw models with leptonic symmetries allow right-handed (RH) neutrino masses at the electroweak scale, or even lower, at the same time having large Yukawa couplings with the Standard Model  leptons, thus yielding observable effects at current or near-future lepton-flavour-violation (LFV) experiments. These models have been previously considered also in connection to low-scale leptogenesis, but the combination of observable LFV and successful leptogenesis has appeared to be difficult to achieve unless the leptonic symmetry is embedded into a larger one. In this paper, instead, we follow a different route and consider a possible connection between large LFV rates and Dark Matter (DM). We present a model in which the same leptonic symmetry responsible for the large Yukawa couplings guarantees the stability of the DM candidate, identified as the lightest of the RH neutrinos. The spontaneous breaking of this symmetry, caused by a Majoron-like field, also provides a mechanism to produce the observed relic density via the decays of the latter. The phenomenological implications of the model are discussed, finding that large LFV rates, observable in the near-future $\mu \to e$ conversion experiments, require the DM mass to be in the keV range. Moreover, the active-neutrino coupling to the Majoron-like scalar field could be probed in future detections of supernova neutrino bursts.

\end{abstract}

\maketitle

\section{Introduction}

Among other problems, the Standard Model (SM) of particle physics lacks an explanation of what is the dark constituent of our Universe, as well as the origin of the tiny neutrino masses. As for the latter, the most popular paradigm is to extend the SM with additional fermions, namely right-handed (RH) neutrinos, whose role is to generate tiny masses for the active  neutrinos,  via the so called seesaw mechanism~\cite{Seesaw}. At the same time, one of the RH neutrinos can play the role of dark matter (DM). The simplest formulation of this is in the type-I seesaw scenario~\cite{Seesaw}, where two of the RH neutrinos are responsible for the active-neutrino masses and mixing, whereas the third one can play the role of warm DM~\cite{Asaka:2005an,keVreview}.

The Majorana masses for the RH neutrinos $N_R$ can, in turn, be generated by the spontaneous breaking of a global $U(1)$ symmetry~\cite{Majoron}. In this so-called \emph{Majoron model} an additional complex scalar field is added to the theory to break $U(1)_l$, thus generating Majorana masses for the RH neutrinos, also entailing an interesting phenomenology coming from the presence of the scalar and pseudo-scalar couplings to both active and sterile neutrinos~\cite{Majoron_pheno,ApostolosMajoron}.

A drawback of the ``vanilla'' seesaw model is that the smallness of the active-neutrino masses requires  the Yukawa couplings of the RH neutrinos to be very small for RH-neutrino masses at reach of current or near-future experiments, thus rendering the model difficult to test in the foreseeable future. The required Yukawa couplings are of order 
\begin{equation}\label{eq:naive_seesaw}
 h_{seesaw} \ \approx \  6 \, \times \, 10^{-8} \, \times \, \sqrt{\frac{m_{\nu_L}}{0.1 \,\mathrm{eV}}} \, \times \, \sqrt{\frac{m_{N_R}}{\mathrm{GeV}}}\;.
 \end{equation}

However, a number of variants of the type-I seesaw mechanism have been developed (e.g the inverse seesaw~\cite{inverseseesaw}, linear seesaw~\cite{linearseesaw}, etc.), where the presence of a leptonic symmetry $U(1)_l$ protects the smallness of the active neutrino masses, thus allowing for much larger Yukawa couplings\footnote{This can be achieved also by means of a discrete symmetry, see e.g.~\cite{Dev:2013oxa}.} than in \eqref{eq:naive_seesaw}, even of order $10^{-3}$ or higher~\cite{largeyukawas}. Therefore, these models provide a way to test the seesaw mechanism in the near future, for instance by the observation of lepton-flavour-violation (LFV) processes, such as $\mu \to e \gamma$ and $\mu \to e$ conversion in nuclei. In particular, the sensitivity of the latter will improve by several orders of magnitude in the near future, thanks to the planned experiments Mu2e and COMET, as well as to the more distant proposal PRISM/PRIME. Typically, in this class of models, in order to generate the observed pattern of neutrino masses and mixing, the leptonic symmetry is explicitly broken \emph{by hand} in different possible ways, giving rise to the so-called inverse-seesaw or linear-seesaw textures, for instance.

Since this class of leptonic symmetries generically gives two quasi-degenerate RH neutrinos, it is tempting to try to explain also the Baryon Asymmetry of the Universe in this model, via the resonant leptogenesis mechanism~\cite{rl, Dev:2014laa}. This can be achieved by supplementing the leptonic symmetry with a larger $O(3)$ symmetry in the RH sector~\cite{Dev:2014laa}. However, it appears to be difficult to reconcile observable LFV rates and successful leptogenesis in the \emph{minimal} models possessing only the leptonic symmetry $U(1)_l$ (see~\cite{Blanchet:2009kk} and Appendix~A of~\cite{Dev:2014laa}). This is true even if one considers GeV-scale leptogenesis mechanisms via RH-neutrino oscillations (see e.g.~\cite{ARSpheno}) or Higgs lepton-number violating decays~\cite{Hambye:2016sby}. Therefore, is it natural to try to address an alternative question: \emph{is it instead possible, in this class of models, to have observable LFV rates and a successful DM candidate?}

In this paper we construct a model achieving this, in which the same leptonic symmetry $U(1)_l$ responsible for (i) light-neutrino masses with large Yukawa couplings, at the same time (ii) stabilizes one of the RH neutrinos, which is therefore a DM candidate. The spontaneous breaking of $U(1)_l$ involves a \emph{Majoron-like} complex scalar field, which in turn (iii) provides a mechanism to generate successfully the DM candidate in the early Universe, together with its keV-scale mass. The charge assignment under $U(1)_l$ needed to achieve this gives, at the same time, a particular pattern for the breaking of the leptonic symmetry, which in our model is \emph{not} performed by hand, but is instead related to the above points.

After this introduction, in section~\ref{sec:model} we will construct the model, derive the mass matrix of the neutrinos after the spontaneous breaking of the electroweak symmetry and $U(1)_l$, and describe quantitatively the generation of DM. In section~\ref{sec:pheno} we study the phenomenology of the model, in particular at near-future $\mu \to e$ conversion experiments, as well as in direct searches of the RH neutrinos. We also study the interactions of the Majoron-like field, which can give detectable imprints at the observation of future supernovae. Finally, in section~\ref{sec:conclusion} we draw our conclusions.

%%%%%%%%%%%%%%%%%%%%%%%%%%%%%%%%%%%%%%%%%%%%%%%%%%%%%%%%%%
\section{The model}\label{sec:model}
As outlined in the introduction, we aim to build a model in which a global $U(1)_l$ leptonic symmetry allows to have a low-scale seesaw mechanism with large Yukawa couplings, and at the same time stabilizes one of the RH neutrinos, identified as a DM candidate.

In the basis in which the RH-neutrino masses are approximately (in a sense that will be made clearer below) real and diagonal, a low-scale seesaw mechanism with large Yukawa couplings is possible if the SM leptons are coupled to the particular combination 
\begin{equation}
N_{+} \equiv \frac{N_2 + i N_3}{\sqrt{2}}\;.
\end{equation} 
As a matter of fact, any arbitrary linear combination can be reduced to this, after rephasing $N_{2,3}$ in order to have their diagonal mass entries real and positive. Therefore, the SM leptons and $N_+$ need to have the same charge under $U(1)_l$, which we take equal to 1, without loss of generality. If the remaining RH neutrino $N_1$ has an even charge under $U(1)_l$, it is absolutely stable to all orders in perturbation theory, since its decay must involve an odd number of neutrinos, assuming that no other scalar fields that develop a vacuum-expectation-value (vev) have a odd charge under $U(1)_l$. Also, its mixing with the remaining fermion fields is forbidden by construction. For simplicity, we fix its charge to 2. Notice also that its nonzero  charge forbids a Majorana mass term, thus making it massless in the symmetric limit of the model. The remaining ingredient is the mechanism responsible for the breaking of $U(1)_l$, which we take as the simplest possible one: a Majoron-like complex scalar field $\Sigma$, charged under $U(1)_l$, that develops a vev. In view of the discussion above, its charge must be even to ensure the stability of $N_1$. As will be shown in the following, the simplest choice $Q_l(\Sigma) = 2$ gives rise to interesting phenomenology.
Thus, the charge assignments of the different fields in our model are summarized in Table~\ref{tab:charges}.
\begin{table}
 \centering
\begin{tabular}{c | c | c | c | c | c }
& $~N_1~$ & $~N_\pm=\frac{N_2\pm iN_3}{\sqrt 2}~$ & $~L~$ &$~e_R~$ & $~\Sigma~$\\
\hline \hline
$Q_l~$& 2 & $\pm 1$&\, +1 \,&\, +1 \,&\, 2\, 
\end{tabular}
\caption{\label{tab:charges} Charge assignment of the different fields under the leptonic symmetry group $U(1)_l$.}
\end{table}

The most general Yukawa and Majorana Lagrangians are 
\begin{align}\label{eq:lagrangianZero}
\mathcal L_Y \ &= \  h_0^l \, \bar L_l  \tilde \Phi  N_+ \,+\, h.c. \;,\notag\\
\mathcal{L}_N \ &= \ - \, 2 \, M_R \, \bar N_+^c N_- \, - \, 2\, g_{++}\, \Sigma^{\dagger}\bar N_+^c N_+ \notag\\
&\quad\; \,-\,  2\, g_{--} \, \Sigma\bar N_-^c N_- \;+\;h.c. \;,
\end{align}
where ${h_0} = (a,b,c)$ and $\tilde \Phi = i \sigma_2 \Phi^*$. Before any spontaneous symmetry breaking, the Yukawa matrix has the form
\begin{equation}\label{eq:h0}
\begin{pmatrix}
0 & a &  i\,a\\
0 & b &  i\,b\\
0 & c &  i\,c\\
\end{pmatrix}\,.
\end{equation}
As mentioned earlier, such structure of the Yukawa matrix  protects the neutrino masses to remain zero at all orders~\cite{Pilaftsis:1991ug}, while the Yukawa couplings can be much larger than in the standard type-I seesaw scenario. Typically, arbitrary perturbations are added to \eqref{eq:h0}, chosen as to fit the neutrino experimental data. Here, instead, the use of a spontaneous breaking of $U(1)_l$ will give a specific pattern for such perturbations in a non trivial manner, due to the particular choice of charge assignment (Table~\ref{tab:charges}).

In the Lagrangian it will also be present, in general, a Higgs-portal coupling $\Phi^\dag \Phi \Sigma^\dag \Sigma$. Its effect in the scalar sector is studied in detail in~\cite{ApostolosMajoron}. Here we assume that its coupling is small enough to not affect significantly Higgs physics. As about the spontaneous breaking of $U(1)_l$ when $\Sigma$ acquires a vev $u$, a particularly interesting scenario is obtained when the Lagrangian mass term for $\Sigma$ is smaller than the other scales in the scalar sector: in this case the phase transition breaking $U(1)_l$ coincides with the electroweak phase transition~\cite{ApostolosMajoron}, and $u_T(T) \propto v_T(T)$, where $T$ is the temperature in the early Universe, $v$ is the Brout-Englert-Higgs (henceforth Higgs for brevity) vev, and the subscript $\!\phantom{.}_T$ denotes the $T$-dependent vevs. 

One has yet to point out a rather generic feature of the model. The Lagrangian\eqref{eq:lagrangianZero}, with $U(1)_l$ broken by the vev of $\Sigma$, is no sufficient for realizing a convenient perturbation term ${\delta h}$ of \eqref{eq:h0}, in the sense explained above. Namely, a mass term for $N_1$ is not generated, and the global rank of the $6 \times 6$ neutrino mass matrix is 3 giving, in addition to one massless RH neutrino, two massless active neutrinos, which is excluded by neutrino oscillation data~\cite{pdg}. In order to circumvent this problem, which is due to the presence of only three ``flavoured'' couplings $(a,b,c)$ in the dimension-4 Lagrangian \eqref{eq:lagrangianZero}, we assume that some UV physics, preserving $U(1)_l$, generates effective operators suppressed by a UV-physics scale $\Lambda$\footnote{Note that such operators would get corrections at the loop level from our initial renormalizable Lagrangian \eqref{eq:lagrangianZero}, but such corrections would have values aligned with the Yukawas $(a,b,c)$ and, as a matter of fact, would not be enough to fit the neutrino oscillation data.}. 
Note that no operator triggering a decay of $N_1$ can be written down, due to our charge assignment, as discussed above.

\subsection{The Lagrangian}

As described above, the most general Lagrangian preserving $U(1)_l$ up to dimension-5 operators can be written as
\begin{equation}\label{eq:lagrangian}
\mathcal L \ = \ \mathcal L_{N} +\mathcal L_{y}+\mathcal L_5 \;,
\end{equation}
where $\mathcal L_{N}$ and $\mathcal L_{y}$ are given by \eqref{eq:lagrangianZero} and 
\begin{equation}\label{eq:lag5}
\mathcal{L}_5 \ = \ -c_{11} \, \frac{\left(\Sigma^{\dagger}\right)^2}{\Lambda} \, \bar N_1^c N_1 \, - \, { c_h^l} \, \frac{\Sigma}{\Lambda} \, N_- (\bar L_l  \tilde \Phi) \, + \, h.c.\,.
\end{equation}
Without loss of generality we can set $c_{11} = 1$, by an appropriate rescaling of $\Lambda$.

After the spontaneous breaking of the global $U(1)_l$ symmetry, writing $\Sigma=u+\frac{S+iJ}{\sqrt 2}$, the Majorana mass matrix of RH neutrinos becomes
\begin{equation}\label{eq:majorana}
\mathcal M_M \ = \ \begin{pmatrix}
\frac{u^2}{\Lambda} &0 & 0 \\
0 & M_R + u \kappa_R& i u \delta_R\\
0&i u \delta_R& M_R-u \kappa_R
\end{pmatrix}\,,
\end{equation}
where we have defined $\kappa_R \equiv g_{++}+g_{--}$ and $\delta_R \equiv g_{++}-g_{--}$. The Dirac mass matrix takes on the form
\begin{equation}\label{eq:mdirac}
\mathcal M_D \ = \ \frac{v}{\sqrt 2}\ ({ h_0} -{ \delta h})\;,
\end{equation}
where ${ h_0}$ is given by \eqref{eq:h0} and
\begin{equation}
{ \delta h} \ = \ \begin{pmatrix}
0 & 0 &  \delta a\\
0 & 0 &  \delta b\\
0 & 0 &  \delta c\\
\end{pmatrix}\,,\qquad \text{~~with~~}\begin{pmatrix}\delta a\\\delta b\\\delta c\end{pmatrix} \ \equiv \  { c_h^l } \, \frac{u}{\Lambda}\;.
\end{equation}

As pointed out above, one ends up with two almost degenerate RH neutrinos,
whereas the lightest one $N_1$ is completely decoupled from the rest of the neutrino sector and constitutes a natural DM candidate, whose mass $\frac{u^2}{\Lambda}$ will be required to be at the $\mathrm{keV}$ scale, as we will see below.

\subsection{Dark Matter Production}
Due to its feeble interactions with the other particles, $N_1$, as a DM candidate, would not be produced thermally in the early Universe. Therefore, its production has to rely either on the annihilation of some interacting particle~\cite{FreezeIn} or through the decay of another particle~\cite{DecayProduction1, DecayProduction2}. The dominant production mechanism in our model is provided by the the decay of the scalar component $S$ of the complex field $\Sigma$, that we will assume to be produced in thermal equilibrium with SM particles after inflation, for instance thanks to its Higgs-portal coupling.

The decay process $S \rightarrow N_1 N_1$ is generated at tree-level once $\Sigma$ acquires a vev $u_T(T)$ (see \eqref{eq:lag5}), which in the early Universe is temperature-dependent. Assuming that $U(1)_l$ is broken during the electroweak phase transition (see the discussion above and~\cite{ApostolosMajoron}), we take $u_T(T) = u \, (1-(T/T_c)^2)^{1/2}$, where $T_c \simeq 160\, \text{GeV}$ is the electroweak critical temperature. Analogously, for the vev-dependent mass of $S$ we take $M_S(T) = m_S \, (1-(T/T_c)^2)^{1/2}$.

\begin{figure}
\includegraphics[width=0.35\textwidth]{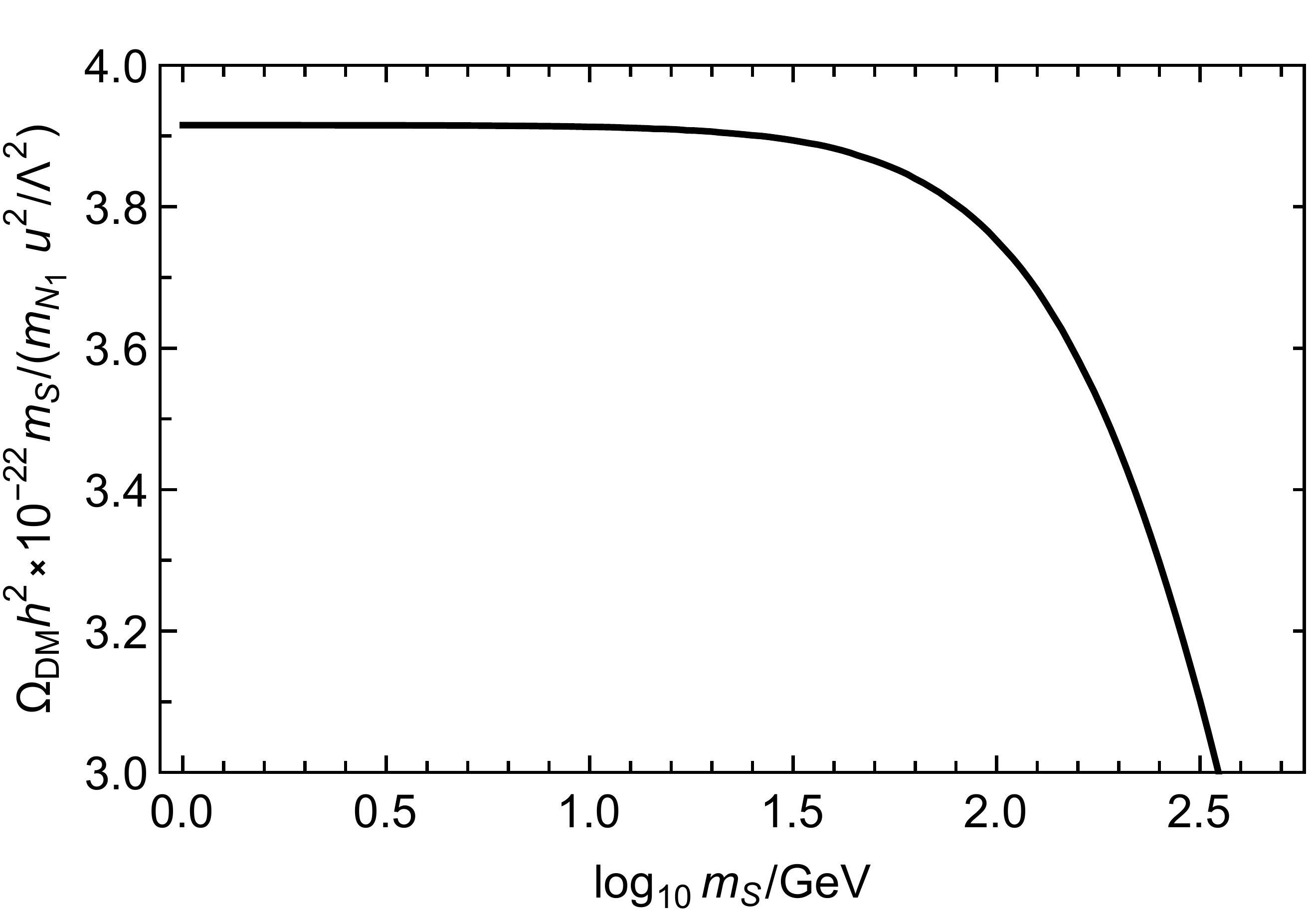}
\caption{\footnotesize The $N_1$ relic abundance produced by the decay mechanism $S \rightarrow N_1 N_1$.\label{fig:dm}}
\end{figure}

Taking into account the decay process $S \rightarrow N_1 N_1$, the Boltzmann equation describing the evolution of the $N_1$ normalized number density $Y(z) \equiv n_{N_1}(z)/s(z)$, with $z \equiv m_S/T$, is
\begin{equation}\label{eq:boltzmann}
z H(z) s(z) \, \frac{d Y}{d z} \ = \ \bigg( 1 - \frac{Y(z)^2}{Y_{\rm eq}(z)^2} \bigg) \, \gamma_D(z) \;,
\end{equation}
where $H(z)$ is the Hubble parameter in the early Universe, $s(z)$ the entropy density, $Y_{\rm eq}(z)$ the equilibrium normalized number density and $\gamma_D(z)$ is the thermally-averaged decay rate. Notice that other decay channels of $S$ to different species do not not contribute to this Boltzmann equation, as long as $S$ is assumed to be in thermal equilibrium. In the freeze-in regime the second term in parentheses in \eqref{eq:boltzmann} can be neglected. The rate $\gamma_D$ is given by
\begin{equation}
\gamma_D(z) \ = \ \frac{m_S^3 \, M_S(z)}{4 \pi^3 z} \; K_1\bigg(\frac{M_S(z)}{T}\bigg) \, \frac{u_T(z)^2}{\Lambda^2} \;.
\end{equation}

In Fig.~\ref{fig:dm} we plot the $N_1$ relic abundance as a function of the zero-temperature mass $m_S$. We see that, for $S$ lighter than about $100 \, \text{GeV}$, we have
\begin{equation}
\Omega_{N_1} h^2 \ \simeq \ 4 \times 10^{22} \,\times \, \frac{u^2}{\Lambda^2} \, \frac{m_{N_1}}{m_S} \;.
\end{equation}
By matching this with the observed relic density $\Omega_{\rm DM} h^2 \simeq 0.12$~\cite{Planck, WMAP} we finally obtain the constraint
\begin{equation}\label{eq:relicdensity}
\frac{u^2}{\Lambda^2} \, \frac{m_{N_1}}{m_S} \ \simeq \ 3 \times 10^{-24}\;.
\end{equation}
Notice that, if the phase transition breaking $U(1)_l$ does not coincide with the electroweak one, but occurs at a higher temperature, this relation will be approximately valid for $m_S$ up to this critical temperature.
%
%Such condition will be used in all the following scans to constrain the scalar field vev $u$ with respect to the UV physics scale $\Lambda$. Note that just requiring non over-abundance of dark matter would relax this constraint to an upper bound on the ratio $u/\Lambda$. However going away from the exact relic abundance will produce less observable signatures. We hence stick to take this condition as an equality, assuming $N_1$ is the only dark constituent of our Universe content. \textcolor{blue}{\it [Lucien : Check that I'm not saying bulsheet here..]}

In addition to the decay process $S \to N_1 N_1$, one can wonder if the scattering $S S \to N_1 N_1$ (in addition to the ones involving $J$) can give a significant contribution to $\Omega_{DM}$. The corresponding thermally-averaged rate is $\gamma_S = T^6/(16 \pi^5 \Lambda^2)$, and the contribution of this process to the relic density is found to be
\begin{equation}\label{eq:relic_scat}
\Omega_{N_1} h^2 \big|_{scat.} \ \simeq \ 2 \times 10^{-14} \times \frac{m_{N_1}}{\text{keV}} \times \bigg(\frac{10^{14} \, \text{GeV}}{\Lambda} \bigg)^2 \times \frac{T_{\rm in}}{\text{GeV}} \;,
\end{equation}
where $T_{\rm in}$ is the initial temperature for the process, i.e. the temperature at which a thermal population of $S$ is generated. The contribution in \eqref{eq:relic_scat} is much smaller than the observed relic density in the parameter space considered below, unless $T_{\rm in}$ is larger than $10^{11} \, \text{GeV}$. Also, UV physics at the scale $\Lambda$ could in principle provide additional mechanisms for the production of $N_1$. However, these are again negligible unless the reheating temperature of the Universe is very high, at a scale comparable to $\Lambda$. Therefore, we will neglect these potential contributions in the following.

After the spontaneous breaking of $U(1)_l$, \eqref{eq:majorana} gives the DM mass:
\begin{equation}\label{eq:dm_mass}
m_{N_1} \ = \ \frac{u^2}{\Lambda} \;.
\end{equation}
Combining \eqref{eq:relicdensity} and \eqref{eq:dm_mass} we can express $m_S$ and $u$ as
\begin{align}
m_S \ &= \ \bigg(\frac{m_{N_1}}{\text{keV}}\bigg)^2 \times \frac{10^{14} \, \text{GeV}}{\Lambda}  \times 3.3 \, \text{MeV} \label{eq:constraint_s}\;,\\
u \ &= \ \sqrt{\frac{m_{N_1}}{\text{keV}}} \times \sqrt{\frac{\Lambda}{10^{14} \, \text{GeV}}} \, \times 10 \, \text{TeV} \;.\label{eq:constraint_u}
\end{align}
These relations will be used below to fix the value of $u$ in the phenomenological scans and to estimate the LFV rates in the next section.

\begin{figure}
\includegraphics[width=0.35\textwidth]{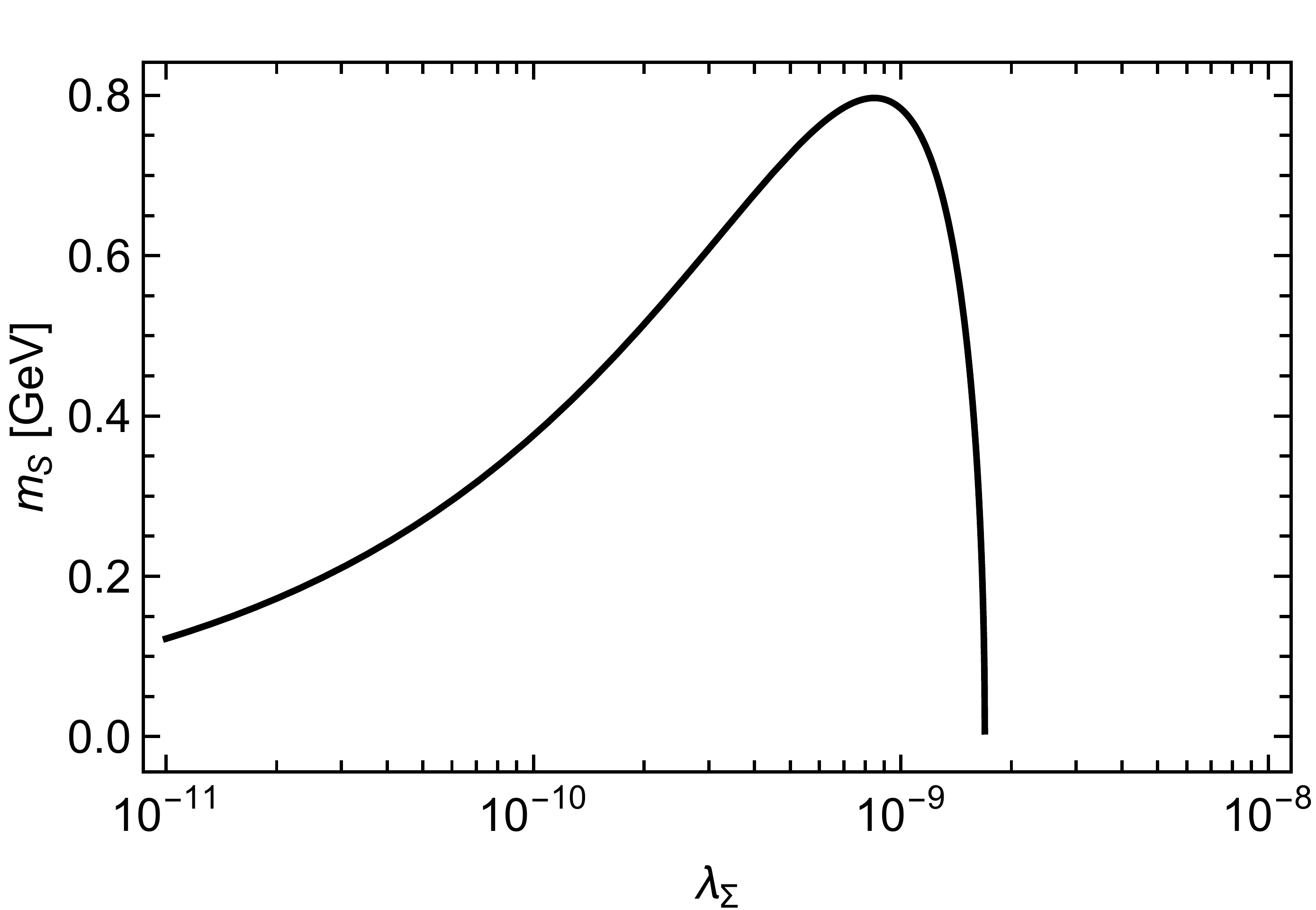}
\caption{\footnotesize The mass of the scalar $S$ as a function of the quartic coupling $\lambda_\Sigma$ for $m_{N_1} = 15 \, \text{keV}$, $\Lambda = 10^{14}\, \text{GeV}$, having imposed the DM constraint~\eqref{eq:constraint_u}.
\label{fig:mS}}
\end{figure}

\subsection{The scalar sector}
The Lagrangian of the scalar sector is
\begin{align}
\mathcal{L}_{\rm scal} \ &= \ \mu^2_\Phi \Phi^\dag \Phi \;+\;\mu^2_\Sigma \Sigma^\dag \Sigma \;+\; \frac{\lambda_\Phi}{2} (\Phi^\dag \Phi )^2 \;+\; \frac{\lambda_\Sigma}{2} (\Sigma^\dag \Sigma )^2 \notag\\ &- \;  \delta_{\Phi\Sigma} \, \Phi^\dag \Phi \, \Sigma^\dag \Sigma \;.
\end{align}
As pointed out in~\cite{ApostolosMajoron} and discussed above, in the regime $|\mu_\Sigma^2| \ll (\delta_{\Phi\Sigma}/\lambda_\Phi) |\mu_\Phi^2|$ the vev $u$ is determined by the electroweak one $v$ as
\begin{equation}\label{eq:vevs}
u \ = \ \sqrt{\frac{\delta_{\Phi\Sigma}}{\lambda_\Sigma}} \; v \;,
\end{equation}
and we define $t_\beta \equiv v/u = \sqrt{\lambda_\Sigma/\delta_{\Phi\Sigma}}$. The mass of $S$ is then given by~\cite{ApostolosMajoron}
\begin{equation}\label{eq:mS}
m_S^2 \ = \ \frac{v^2}{2} \bigg[ \, \lambda_\Phi + \lambda_\Sigma t_\beta^{-2} - \sqrt{\big(\lambda_\Phi - \lambda_\Sigma t_\beta^{-2}\big)^2 + \delta_{\Phi\Sigma}^2 t_\beta^{-2}}\,\bigg]\;.
\end{equation}

We now show that the conditions \eqref{eq:constraint_s} and \eqref{eq:constraint_u}, fixed by the DM mass and observed relic density, can be successfully realized in the interesting region of the parameter space, with natural choices for the parameters of the scalar sector. 

Let us first consider a benchmark scenario with $m_{N_1} = 15 \, \text{keV}$, $\Lambda = 10^{14}\, \text{GeV}$ which, as we will show in the next section, is in the region where the model predicts interesting phenomenology.  By using \eqref{eq:constraint_u} and \eqref{eq:vevs} in \eqref{eq:mS}, we may find $m_S$ as a function of $\lambda_\Sigma$, which we plot in Fig.~\ref{fig:mS}.
For the benchmark point that we are considering, the DM constraint \eqref{eq:constraint_s} gives $m_S \approx 0.75 \, \text{GeV}$; Fig.~\ref{fig:mS} shows that this can be realized, without fine-tuning, for $\lambda_\Sigma \approx 10^{-9}$, and consequently $\delta_{\Phi\Sigma} \approx 2 \times 10^{-5}$.

One may wonder if such small values for the couplings are natural, in the sense that their loop corrections are smaller than themselves. It is easy to convince oneself that this is the case, at least at 1-loop order: for instance, the most dangerous correction to $\lambda_\Sigma$ comes from the diagram with four external $\Sigma$ legs and a Higgs loop which, up to logarithms, scales as $\delta_{\Phi\Sigma}^2/(16 \pi^2) \sim 10^{-12} \ll \lambda_\Sigma \sim 10^{-9}$.

However, there are regions of the parameter space where one has large LFV rates without having such small couplings in the scalar sector. To show this, let us consider a second benchmark scenario with $m_{N_1} = 100 \, \text{keV}$, $\Lambda = 10^{13}\, \text{GeV}$ and Wilson coefficients $c_h \sim 0.1$. The estimate \eqref{eq:xi} in the next section shows that the LFV rates are large, for these values of the parameters. On the other hand, \eqref{eq:constraint_s} gives $m_S \simeq 300\, \text{GeV}$, which in turn involves couplings in the scalar sector of the order of the SM quartic one.

\section{Phenomenology}\label{sec:pheno}

\subsection{Lepton Flavour Violation}

The light-neutrino mass matrix is given by the seesaw formula:
\begin{equation}\label{eq:seesaw}
m_\nu \ \simeq \ - \mathcal{M}_D \cdot \mathcal{M}_M^{-1} \cdot \mathcal{M}_D^{\rm T} \;,
\end{equation}
with $\mathcal{M}_D$ and $\mathcal{M}_M$ given by \eqref{eq:mdirac} and \eqref{eq:majorana}, respectively.
For $m_\nu$ we adopt the standard parametrization of the PMNS matrix~\cite{pdg}, and fix the mass differences and mixing angles to their best-fit values~\cite{pdg}. Because of the leptonic symmetry $U(1)_l$, only two RH neutrinos effectively participate in the seesaw mechanism, and therefore the mass of the lightest neutrino vanishes in our model. In the following, we will restrict to the inverted-hierarchy spectrum of active neutrinos, because this gives larger LFV rates involving the $e$ and $\mu$ flavours, which is the main focus of this paper. Thus, for a particular choice of the Dirac and Majorana phases $\delta$ and $\phi_{1,2}$, the light-neutrino mass matrix is completely determined, and \eqref{eq:seesaw} gives 5 independent (complex) relations to fix a number of model parameters. In particular, in addition to the PMNS phases $\delta,\phi_{1,2}$, we take as input parameters for the seesaw relation\footnote{Note that the terms involving $g_{++}$ cancel in the seesaw formula at leading order.}: $\Lambda, M_R, c_h^a, g_{--}$ and $u$, as determined by \eqref{eq:constraint_u}. Then, \eqref{eq:seesaw} is used to determine $a, b, c, c_h^b, c_h^c$.

\begin{figure}
\includegraphics[width=0.45\textwidth]{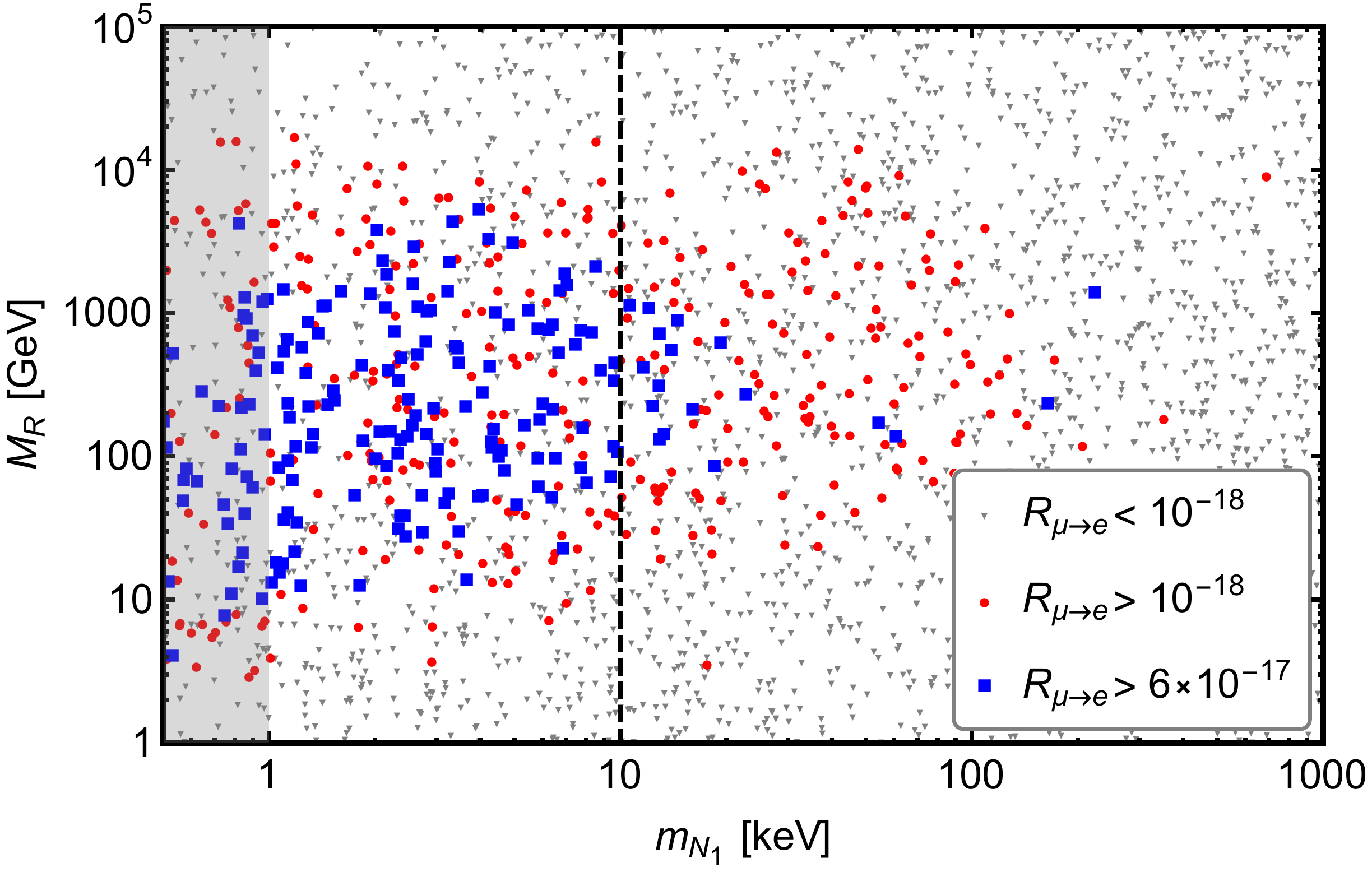}
\caption{\footnotesize Results of the numerical scan with $\Lambda = 10^{14}\,\text{GeV}$, $c_h^a = 1$. The shaded region $m_{N_1} < 1\, \text{keV}$ corresponds to the region excluded the Tremaine-Gunn bound, the dashed line to the structure-formation bound \eqref{eq:lyman1}.
\label{fig:scan1}}
\end{figure}

Recalling that in the $U(1)_l$ symmetric limit the light-neutrino mass matrix vanishes, there are two leading contributions to $m_\nu$ from \eqref{eq:seesaw}, one coming from the $U(1)_l$ breaking in $\mathcal{M}_D$ (the linear-seesaw term) and one from the breaking in $\mathcal{M}_M$ (the inverse-seesaw term). Respectively, these are schematically given by
\begin{align}
m_{\nu} \ &\supset \ \frac{v^2 \, h_0}{M_R} \, c_h \, \frac{u}{\Lambda} \;,\label{eq:cont1}\\
m_{\nu} \ &\supset \ \frac{v^2 \, h_0^2}{M_R^2} \, 2 \, u \, g_{--} \;,\label{eq:cont2}
\end{align}
where $h_0$ and $c_h$ denote, collectively, the entries of the corresponding vectors. Barring the possibility of fine-tuning, i.e.~of large cancellations between these two terms in the seesaw relation, they need to be separately of the order of the atmospheric mass difference\footnote{For the term coming from the breaking in $\mathcal{M}_M$ this is actually an upper bound, since it is possible to fit the light-neutrino mass matrix with the $U(1)_l$ breaking term in $\mathcal{M}_D$ only.} $\Delta m_{\rm atm} \simeq 49 \, \text{meV}$. Using also \eqref{eq:constraint_u} we thus obtain the order-of-magnitude estimates:
\begin{align}
c_h^l \, \xi \ &\approx \ 1.5 \times 10^{-3}\, \times \sqrt{\frac{\Lambda}{10^{14} \, \text{GeV}}} \,\times \, \sqrt{\frac{\text{keV}}{m_{N_1}}} \;,\label{eq:xi}\\
g_{--} \, \xi^2 \ &\lessapprox \ 1.3 \times 10^{-15}\, \times \sqrt{\frac{10^{14} \, \text{GeV}}{\Lambda}} \,\times \, \sqrt{\frac{\text{keV}}{m_{N_1}}} \label{eq:gmm}\;,
\end{align}
where $\xi$ is the light-heavy neutrino mixing parameter $\xi \equiv h_0 v / (\sqrt{2} M_R)$. Eq. \eqref{eq:xi} is the central result of this discussion: assuming that the Wilson coefficients $c_h^l$ are $\mathcal{O}(1)$, i.e. that there is no large hierarchy of scales in the UV-completion of the model, in order to have observable LFV effects, which require a mixing $\xi > 10^{-4}$, the scale $\Lambda$ has to be at least $10^{13} \, \text{GeV}$ and the DM mass has to be in the keV range. Note that the parameter-space region where $m_{N_1} \lessapprox 1 \mathrm{keV}$ is excluded by the Tremaine-Gunn bound~\cite{Tremaine:1979we}, namely requiring that a fermionic DM population cannot reach a higher density than the one established by Pauli exclusion principle.

\begin{figure}
\includegraphics[width=0.45\textwidth]{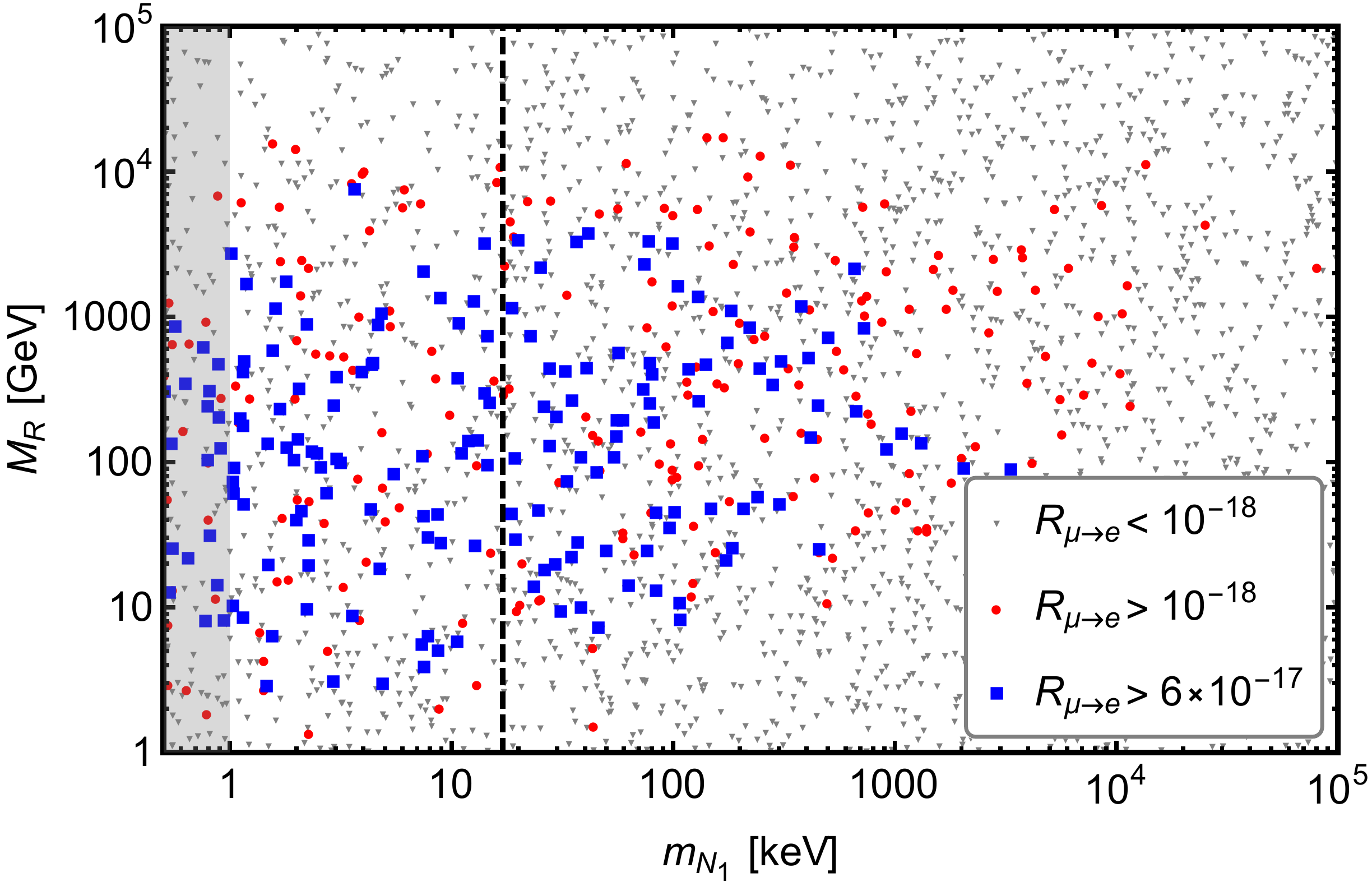}
\caption{\footnotesize Same as Fig.~\ref{fig:scan1}, but with $\Lambda = 10^{16}\,\text{GeV}$, $c_h^a = 1$. The dashed line is the structure-formation bound, as given by \eqref{eq:lyman2}.
\label{fig:scan2}}
\end{figure}

A stronger bound is obtained by structure formation. For non-resonant thermal production of DM, the analysis of Lyman-$\alpha$ forest data~\cite{Viel:2013apy} give the bound on the mass of the DM, as a thermal relic, of $3.3\,\text{keV}$, which translate into a bound on non-resonant thermal production of $m_{\rm NRP} > 22\, \text{keV}$. This quantity can be related to one for the production of DM by decay of an equilibrated particle, of interest here, as~\cite{Bezrukov:2014nza}
\begin{equation}
m_{decay} \ \simeq \ m_{\rm NRP} \, \times \frac{2.45}{3.15} \, \bigg( \frac{10.75}{g_{*}(m_S/3)}\bigg)^{1/3} \;,
\end{equation}
where $g_*(T)$ is the effective number of relativistic degrees of freedom. By using~\eqref{eq:constraint_s} we obtain the bounds
\begin{align}
m_{N_1} \ &\gtrapprox \ 10\,\text{keV}\;, \quad &(\Lambda = 10^{14} \, \text{GeV}) \label{eq:lyman1}\\
m_{N_1} \ &\gtrapprox \ 17\,\text{keV}\;. \quad &(\Lambda = 10^{16} \, \text{GeV}) \label{eq:lyman2}
\end{align}
Notice, however, that these analytic estimates may receive significant corrections in some regions of the parameter space~\cite{Merle:2015oja}.

It is interesting to note here that asking for large LFV rates, as just argued above, provides a naturally high value for the scale $\Lambda$, say at the scale of the intermediate breaking of grand unfication theories (GUTs) ($\sim 10^{13}\,\mathrm{GeV}$) or even at the GUT scale itself. Such intermediate-scale physics, where intermediate subgroups of GUTs break down to the SM, is argued to cure the metastability of the Higgs vacuum in the context of non supersymmetric theories~\cite{HiggsInstability, Higgs2} and is naturally embedded in SO(10) GUTs~\cite{GUT, IntermediateGUT}.

\begin{figure}
\includegraphics[width=0.45\textwidth]{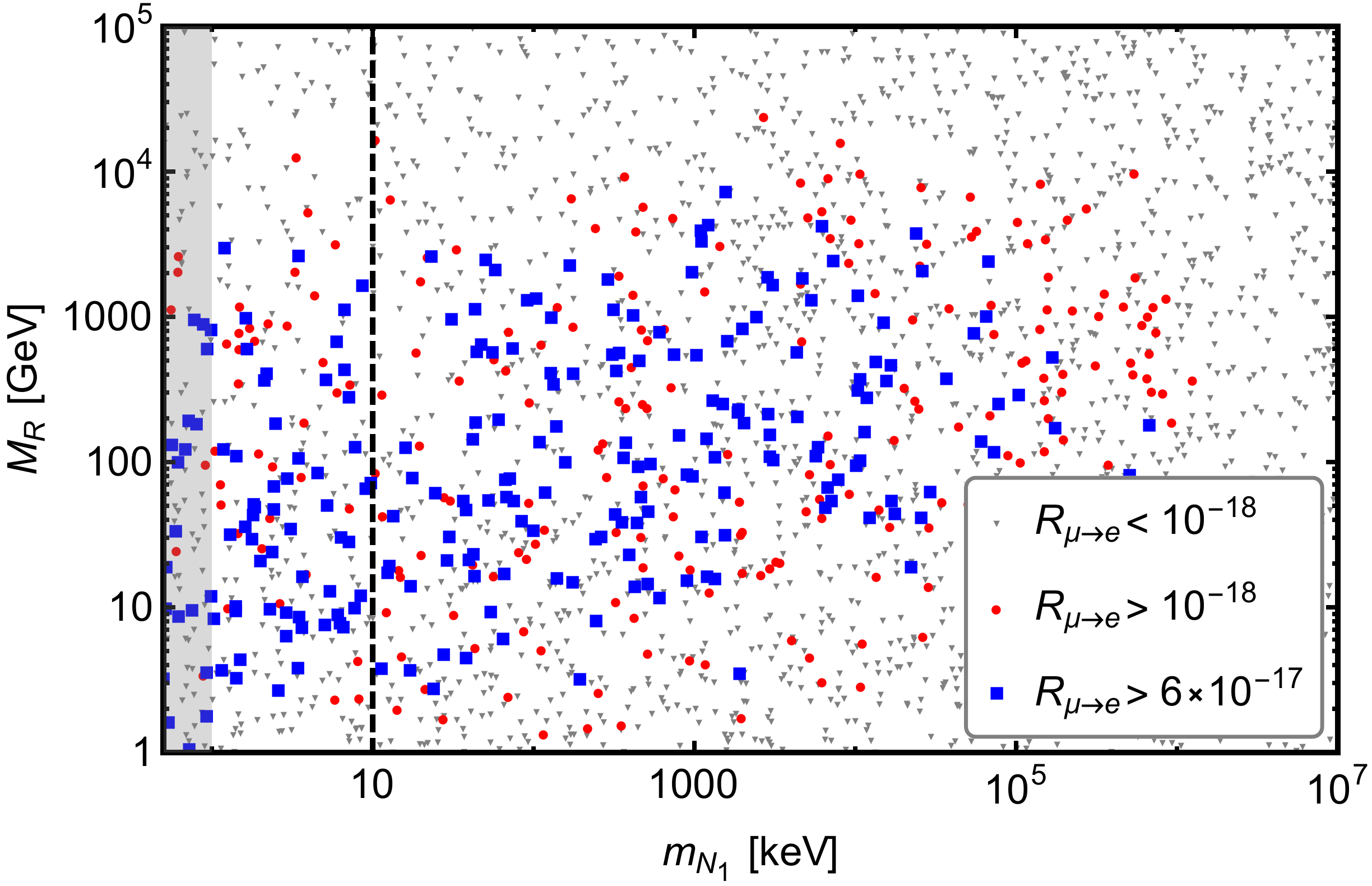}
\caption{\footnotesize Same as Fig.~\ref{fig:scan1}, but with $\Lambda = 10^{14}\,\text{GeV}$, $c_h^a = 0.01$. 
\label{fig:scan3}}
\end{figure}

In order to investigate the LFV phenomenology more quantitatively, we have performed numerical scans of the relevant parameter space, as we are going to describe now. For fixed values of $\Lambda$ and $c_h^a$ we have scanned over the parameters $M_R, m_{N_1}, g_{--}, g_{++}, \phi_{1,2}$, having instead fixed the value of the Dirac phase $\delta = -\pi/2$ (as very mildly suggested by the current oscillation data). The phases $\phi_{1,2}$ and the argument of the complex parameters $g_{--,++}$ are scanned with uniform probability over the range $[0,2\pi[$. The logarithm base 10 of $M_R, m_{N_1}, |g_{--,++}|$ is scanned uniformly too. The latter between the values $-12$ and $-5$. The remaining parameters are obtained as described above. To avoid fine-tuned solutions, we require that there are not large cancellations in the seesaw relation between the two contributions \eqref{eq:cont1} and \eqref{eq:cont2}, in particular that the individual contributions are less than a factor of 10 larger than the overall one.

We calculate the $\mu \to e$ conversion rate $R_{\mu \to e}$ for Aluminium nuclei following~\cite{mutoerates}, and divide the scanned points into three subsets: (i) the ones with $R_{\mu \to e} > 6 \times 10^{-17}$, which will be probed in the near future by the Mu2e~\cite{mu2e} and COMET~\cite{comet} experiments; (ii) the ones with $6 \times 10^{-17} > R_{\mu \to e} > 10^{-18}$, which may be probed in the more distant future by the PRISM/PRIME proposal~\cite{prism}; (iii) the ones with small LFV rates $R_{\mu \to e} < 10^{-18}$. We present our results in Fig.\ref{fig:scan1}--\ref{fig:scan3}, for different values of $\Lambda$ and $c_h^a$. We see that, generically, in order to have observable LFV rates the DM mass has to be in the keV range, whereas the two heavier RH neutrinos can be in the 1 GeV -- 10 TeV range. For values of the Wilson coefficient $c_h^a$ significantly smaller than 1, the DM candidate can also be heavier (see Fig.~\ref{fig:scan3}).

\begin{figure}
\includegraphics[width=0.45\textwidth]{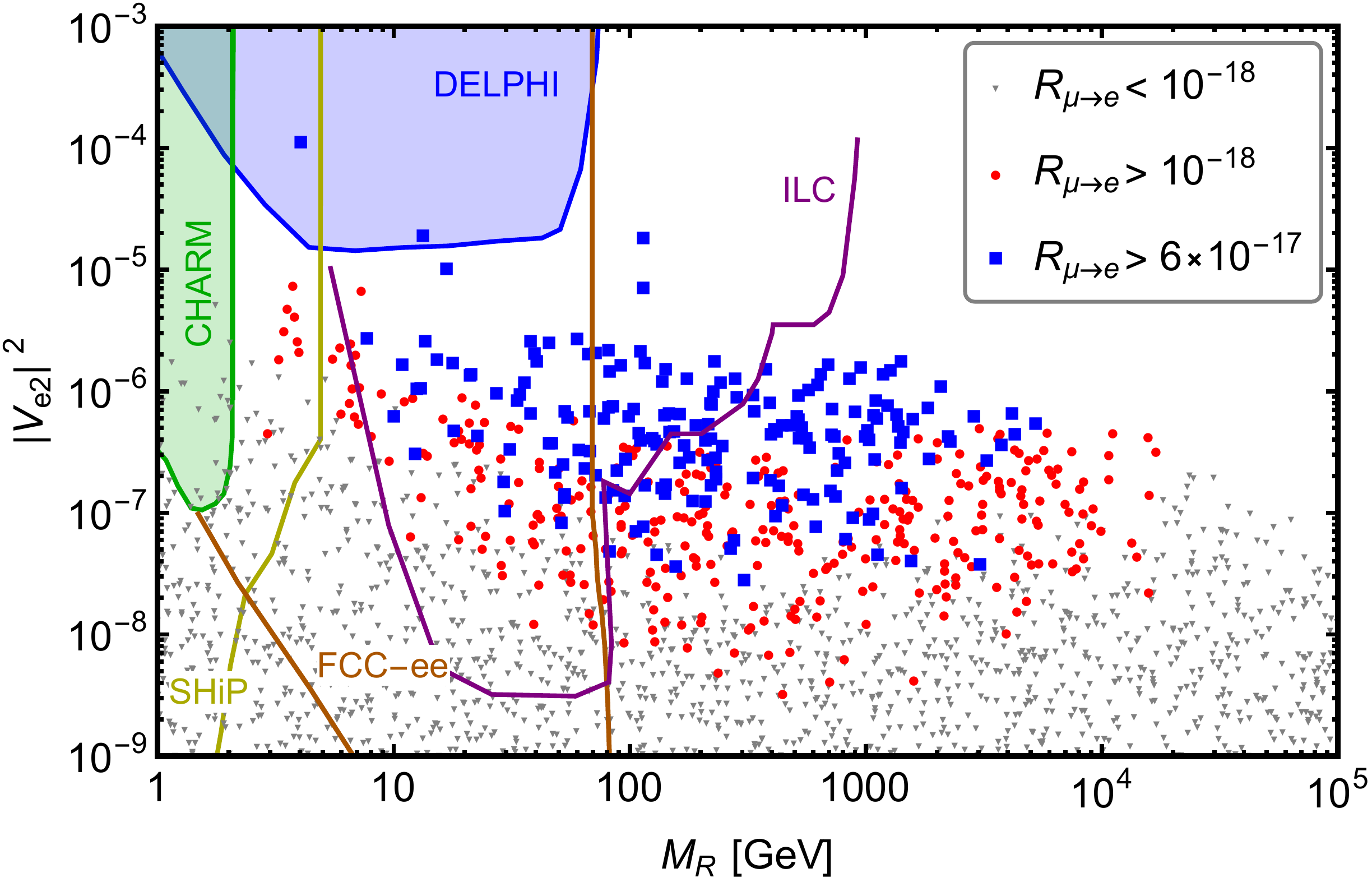}
\caption{\footnotesize Results of the numerical scan with $\Lambda = 10^{14}\,\text{GeV}$, $c_h^a = 1$, as a function of the mass of the heavier RH neutrinos and their mixing with the active electron flavour. Existing bounds and future prospects for direct searches of $N_{2,3}$ are also shown.
\label{fig:direct}}
\end{figure}

In Fig.~\ref{fig:direct} we show the results of the scan as function of $M_R$ and the light-heavy neutrino mixing. We exhibit the relevant existing bounds coming from direct searches of the two heavier RH neutrinos with mass $M_R$, as taken from~\cite{Deppisch:2015qwa}. We also plot the sensitivity of the planned SHiP experiment at CERN~\cite{ship} and the proposed future circular collider FCC-ee~\cite{FCC}, in its run at the Z-pole mass, and preliminary results for the combined sensitivity at ILC~\cite{ILC}. Fig.~\ref{fig:direct} shows that direct searches will also probe a significant region of the parameter space of the model, where large LFV rates are obtained. However, the region with $M_R > 350 \, \text{GeV}$ appears to be probable, in the foreseeable future, only by LFV experiments.

\begin{figure}
\includegraphics[width=0.45\textwidth]{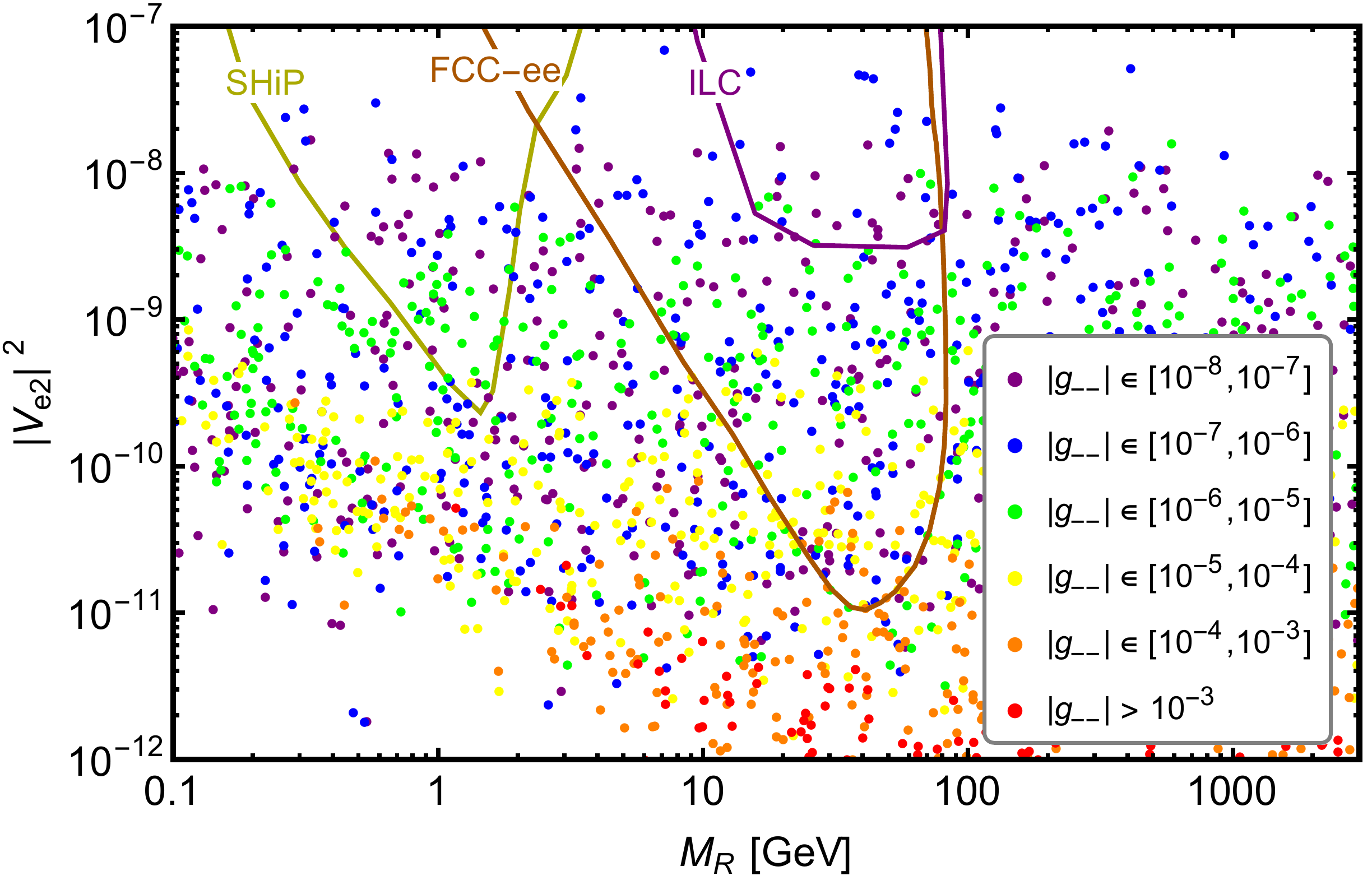}
\caption{\footnotesize Results of the numerical scan with $\Lambda = 10^{11}\,\text{GeV}$, $c_h^a = 0.1$, as a function of the mass of the heavier RH neutrinos and their mixing with the active electron flavour, for different values of $|g_{--}|$. Future prospects for direct searches of $N_{2,3}$ are also shown.
\label{fig:larger_g}}
\end{figure}

Finally, we point out that the estimate \eqref{eq:gmm} gives $g_{--} \lessapprox 10^{-7}$, in the parameter region with observable LFV rates, thus implying a small hierarchy with the Yukawa couplings $h_0 \sim 10^{-5}-10^{-4}$, for $m_N = 10-100\, \text{GeV}$ and $\xi \sim 10^{-4}$. However, larger values of $g_{--}$, of the same order as $h_0$, gives a mixing $\xi$ still in the SHiP and FCC-ee range, as shown in Fig.~\ref{fig:larger_g}.

\subsection{(Pseudo)Scalar interactions}

Now that we have investigated possible signatures at LFV processes, we aim to look for potential cross-signals involving the new scalar interactions with the SM leptons.
Indeed, the coupling of the complex field $\Sigma$ to both right- and left-handed neutrinos might have detectable signatures in leptonic BSM physics searches as well as astrophysics observations.

Let us first mention that, following~\cite{ApostolosMajoron}, loop processes generating decays of the kind $l^- \rightarrow l'^- J$ have been checked to be small in the region of parameter space considered in this paper and therefore are not relevant for our discussion.

As far as neutrino secret interactions are concerned, there have been a number of studies  constraining the emission of Majorons by supernova bursts of neutrinos~\cite{SN1987}. Indeed, a too high production of pseudo scalars out of the supernova core can affect significantly the flux of energy emitted by these objects and hence it gets constrained by the observation of SN1987a. In addition to the presence of the Majoron, a massive scalar is present in the theory and could also be produced copiously by supernova neutrino bursts. The preliminary analysis in~\cite{Yongchao} provides bounds on such massive-particle emission and prospects on possible future SN detections in the next decades, which are particularly relevant in our model. We will thus see how our parameter space is constrained by this observable and up to which point observable LFV effects are compatible to possible signals at future supernova observations.

%\paragraph{\bf Explicit Couplings}
Let us rotate the mass matrix of the whole neutrino sector
\begin{equation}
\mathcal M \ = \ \begin{pmatrix}
0 &\mathcal M_D\\ \mathcal M_D^T& \mathcal{M}_M
\end{pmatrix} \;,
\end{equation}
 into a block-diagonal form by means of the unitary matrix
\begin{equation}
V \ = \ \begin{pmatrix}
(\mathbf 1_3+\xi^\star \xi^T)^{-\frac{1}{2}} & \xi^\star(\mathbf 1_3+\xi^\star \xi^T)^{-\frac{1}{2}}\\
-\xi^T(\mathbf 1_3+\xi^\star \xi^T)^{-\frac{1}{2}}&(\mathbf 1_3+\xi^\star \xi^T)^{-\frac{1}{2}}
\end{pmatrix}\;,
\end{equation}
where we have introduced the matrix $\xi \equiv m_D \mathcal M_N^{-1}$.
%allows to define as a converging series $(\mathbf 1_3+\xi^\star \xi^T)^{-1/2}$ only if $||\xi||=\sqrt{\mathrm{Tr}[\xi^\dagger\xi]}<1$.
Going to this block-diagonal flavour basis one defines
\begin{equation}
\begin{pmatrix}
\widehat \nu_L\\ \widehat N_R
\end{pmatrix} \ \equiv \  V^\dagger\begin{pmatrix}
\nu_L\\ N_R
\end{pmatrix} \ \simeq \ \begin{pmatrix}
\mathbf{1}_3&-\xi^\star\\
\xi^T&\mathbf{1}_3
\end{pmatrix}\begin{pmatrix}
\nu_L\\ N_R
\end{pmatrix}\,.
\end{equation}
Writing the Lagrangian in the flavour basis we can now obtain explicitly the couplings between the active neutrinos and the real component of $\Sigma = u+\frac{s+iJ}{\sqrt 2}$. Here, the Majoron is the pseudo-scalar component $J$, which is massless up to possible quantum-gravity effects~\footnote{Note that in this case the Majoron itself can play the role of a DM candidate, see e.g.~\cite{Majoron_DM}}, while the scalar component $S$ is massive. The couplings to neutrinos read
\begin{equation}
-\mathcal L \ \supset \ \frac{s}{2} \, (\bar\nu_L^c)_{i}(Y^s)_{ij}(\nu_L)_{j} \;+\; \frac{iJ}{2}\, (\bar\nu_L^c)_{i}(Y^J)_{ij}(\nu_L)_{j}\;,
\end{equation}
where the matrix $Y^J$ and $Y^s$ are defined as
\begin{equation}
Y^J \ = \ -\sqrt{2} \; \xi \cdot\begin{pmatrix}
 \frac{u}{\Lambda}&0&0\\
0&\kappa_R&i\delta_R\\
0&i\delta_R&-\kappa_R
\end{pmatrix}\cdot\xi^T\;,
\end{equation}
and
\begin{equation}
Y^s \ = \ \sqrt{2} \; \xi \cdot\begin{pmatrix}
 \frac{u}{\Lambda}&0&0\\
0&\delta_R&i\kappa_R\\
0&i\kappa_R&-\delta_R
\end{pmatrix}\cdot\xi^T \;.
\end{equation}
The coupling to the electron neutrino (the most constrained by astrophysical observations) is the component (1,1) of the these matrices. We find
\begin{align}
(Y^J)_{ee} \ &= \ \frac{-2 \sqrt 2\ a^2 \,g_{--}v^2}{M_R^2-4 g_{--}g_{++}u^2} \  \approx \  4 \sqrt{2} \, \xi^2 \, g_{--}\;,\\
(Y^s)_{ee} \ &=\ \frac{-2\sqrt 2\ a^2 \, g_{--}v^2\left(M_R^2+4g_{--}g_{++}u^2\right)}{\left(M_R^2-4 g_{--}g_{++}u^2\right)^2} \notag\\
\ &\approx \ 4 \sqrt{2} \, \xi^2 \, g_{--}\;.
\end{align}

\begin{figure}
\includegraphics[width=0.45\textwidth]{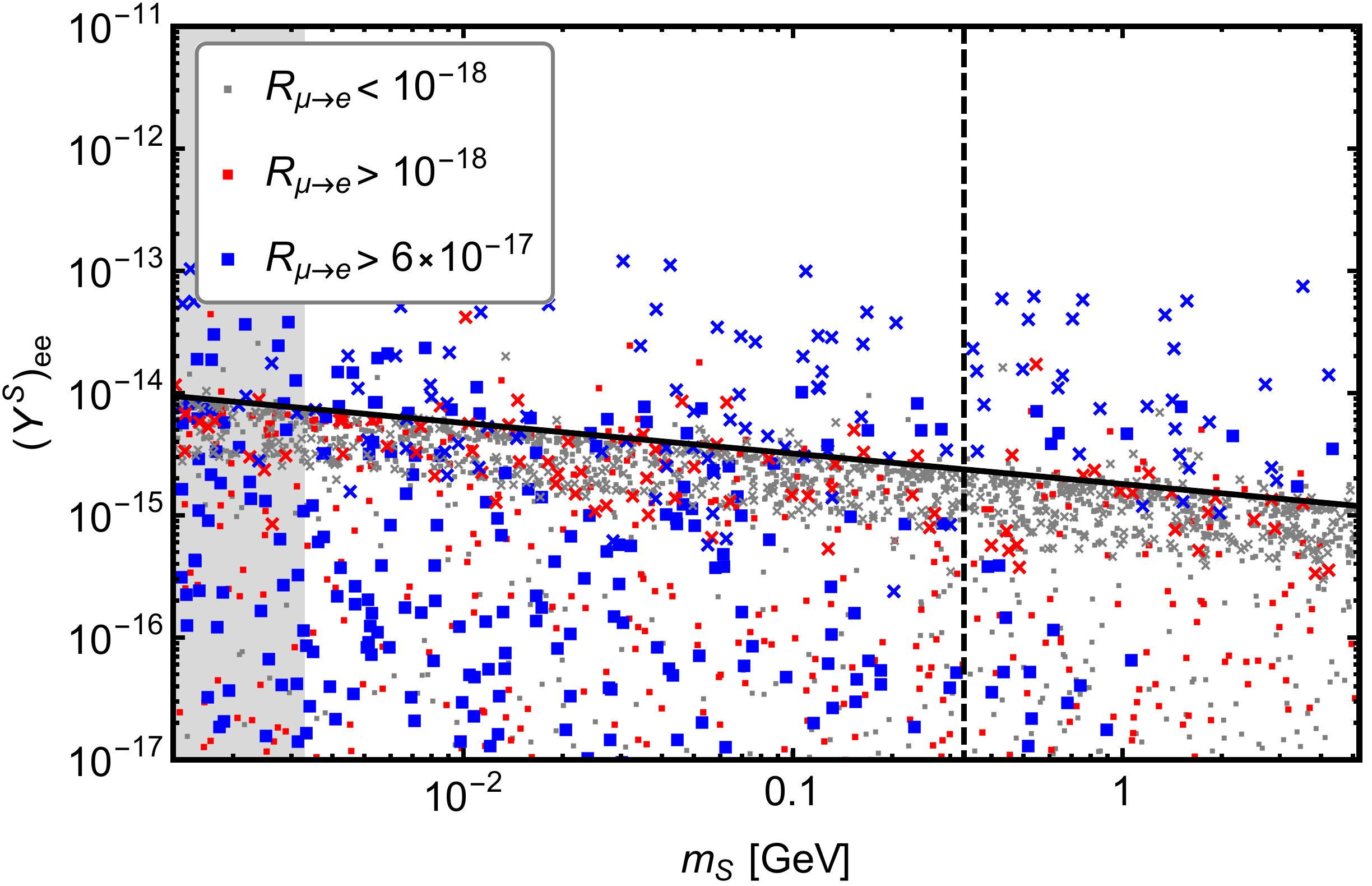}
\caption{\footnotesize The coupling of the scalar to electron neutrinos as a function of $m_S$. Same color code as in Fig.\ref{fig:scan1}, with $\Lambda = 10^{14}\,\text{GeV}$, $c_h^a = 1$. In this plot we also allow for solutions with a moderately larger fine-tuning cut of 100, denoted by crosses. The black line represent the order-of-magnitude bound for such coupling, obtained from \eqref{eq:gmm}, which is released by allowing for a moderate fine tuning. Large LFV rates are compatible with a coupling $(Y^S)_{ee} \gtrsim 10^{-13}$ at $m_S \sim 100\, \text{MeV}$, which is close to the current astrophysical limit \eqref{eq:astro}.
\label{fig:majo}}
\end{figure}

A bound on Majoron emission from the SN1987a supernova core explosion has been derived\footnote{See also~\cite{Forastieri:2015paa}.} in~\cite{SN1987}, excluding the region
\begin{equation}
3\times 10^{-7} \ < \ |(Y^J)_{ee}| \ <  \ 2\times 10^{-5}\;.
\end{equation}
As a matter of fact, the region of parameter space of interest in this paper is far from this exclusion region. Indeed, the order-of-magnitude estimate \eqref{eq:gmm} shows that such coupling is at most of order $10^{-14}$, when the LFV rates are in the observable range. Nevertheless, the study~\cite{Yongchao} of $\mathcal{O}(100)\,\text{MeV}$ particle emission from supernovae imposes much stronger constraints on processes $\nu \nu\rightarrow s$ from supernova energy loss. Indeed, the typical temperature of the neutrino bath present in the supernova cores (constituted mainly of electron neutrinos) is $\mathcal O(10)\,\mathrm{MeV}$, whereas the chemical potential of the lightest species is $\mathcal O(100)\,\mathrm{MeV}$. The potential supernova energy loss by emission of $m_S \, \sim 100\,\mathrm{MeV}$ scalars can thus be strongly constrained by the measurement of SN1987a~\cite{SN1987} leading to the following exclusion region~\cite{Yongchao}
\begin{equation}\label{eq:astro}
2.1\times 10^{-11} \ < \ |(Y^S)_{ee}| \ < \ 1.6\times 10^{-8}\,.
\end{equation}
As shown in Fig.~\ref{fig:majo}, most of the points in the parameter space of the model are outside this region and therefore the requirement of large LFV rates is compatible with the observation of 1987a.

However, as discussed in~\cite{Yongchao}, the bound \eqref{eq:astro} will be improved considerably by the possible detection of future supernova explosions by IceCube, SuperKamiokande, as well as future DM direct-detection experiments~\cite{Lang:2016zhv, Fischer:2016cyd}. In particular, the possible detection of a close-by supernova explosion ($d\lesssim 10\,\mathrm{kpc}$), together with the very accurate measurement  of their neutrino luminosity curves, may improve these bounds by several orders of magnitude, thus giving detectable signals in both LFV experiments and supernova explosions. Namely, the region of the parameter space probed by such a detection would be~\cite{Yongchao} 
\begin{equation}\label{eq:astro2}
2.1\times 10^{-13} \ < \ |(Y^S)_{ee}| \ < \ 2.5\times 10^{-8}\,.
\end{equation}
As seen in Fig.~\ref{fig:majo}, allowing for some fine-tuning in the seesaw relation, part of the points in the parameter space could be  detectable by both LFV experiment and supernova explosion measurements, thus providing a possible double signature for the model.

\section{Conclusion}\label{sec:conclusion}

In this paper, we have investigated the possibility that a leptonic global symmetry $U(1)_l$ protecting the lightness of the active neutrinos in the type-I seesaw scenario, at the same time allowing for sizeable Yukawa couplings with the RH sector, is broken spontaneously by adding one ``Majoron-like'' complex scalar field to the seesaw Lagrangian. The charge assignment under $U(1)_l$ of such setup allows to render one of the RH neutrino absolutely stable, which therefore constitutes a natural DM candidate. Whereas in this framework it would be difficult to produce the DM candidate via its (tiny) interactions with the active sector, the decays of the field responsible for the breaking of $U(1)_l$ allow for a successful production of DM in the early Universe, via a freeze-in mechanism.

In the dimension-4 Lagrangian only 3 ``flavoured'' couplings are present. In order to obtain a sufficient number of couplings to fit the non-trivial active-neutrino mass matrix, dimension-5 effective operators need to be considered too. These can arise from some $U(1)_l$ invariant heavy sector at an intermediate scale $\Lambda$. 

Interestingly, the requirement of having large LFV rates, observable in the near future, as well as the observed DM relic density, fixes the various scales of the model:
\begin{itemize}\itemsep0.3em
\item[$(i)$] the DM mass has to be in $\mathrm{keV}$ range, with possibilities up to the $\mathrm{MeV}$ range, in some regions of the parameter space;
\item[$(ii)$]the scale $\Lambda$ has to be at least $10^{13}\, \mathrm{GeV}$ for $c_h \simeq 1$, which coincides with the scale of intermediate breaking of various GUT models, or with the GUT scale itself;
\item[$(iii)$] the scale of $U(1)_l$ breaking is typically in the $10-1000\,\mathrm{TeV}$ range. As pointed out in~\cite{ApostolosMajoron}, the phase transition breaking $U(1)_l$ can even coincide with the electroweak one.
\end{itemize}
The model -- in addition to be as minimal as possible -- has a set of features which make it testable by future neutrino-physics measurements. A first point is that the requirement of large LFV rates is more easily satisfied for an inverted-hierarchy mass spectrum, although there are possibilities even for a normal-hierarchy spectrum too. More importantly, since only two RH neutrinos have an active role in the seesaw mechanism, the lightest of the active neutrinos is automatically massless. 

By construction, the presence of large Yukawa couplings allows for LFV processes with large rates, detectable in the near-future at $\mu \to e$ conversion experiments Mu2e and COMET. As we have explained above, this requirement, which is the original motivation for the model, fixes its mass scales. In addition to this, since the heavier RH states have masses lighter than 300 GeV in a good portion of the parameter space, the model can be tested also by the direct production of these states at future proposed experiments, such as SHiP, FCC-ee and ILC. Notice that this feature is precisely what distinguishes the model from a ``standard'' Majoron one, since the latter cannot yield a sufficiently large mixing with the SM leptons, unless a severe fine-tuning in the seesaw relation is invoked. Notice also that, along the same lines of the model discussed here, one could construct models with a larger number of fields and where the neutrino mass generation is, for instance, only of the linear-seesaw or inverse-seesaw type, cf.~\eqref{eq:cont1} and \eqref{eq:cont2}. Their phenomenology would be similar to the one discussed here, apart from the number of states in the GeV-TeV range, which needs to be larger in the inverse-seesaw case, and the suppression of the couplings of the scalar to SM leptons, in the linear-seesaw case.

Finally, the coupling of the Majoron-like scalar field to the active neutrinos can be large in a region of the parameter space with $\Lambda \sim 10^{13-14}\, \text{GeV}$, being in particular close to the recent bound obtained from the neutrino burst of supernovae. Therefore, the model can have addition complementary signatures at future supernova detections by IceCube and SuperKamiokande, which would provide an additional strong piece of evidence for it.

\medskip

\section*{Acknowledgements}
The authors would like to thank Y.~Zhang and J.~Heeck for interesting discussions, and Alexander Merle for useful comments on the structure-formation bound. L.H. would like to thank the DESY theory group of Hamburg for its hospitality during the final stage of preparation of this work, as well as Sophie Martin for interesting and lively discussions.
The work of L.H. has been partly funded by PIER (Partnership for Innovation, Education and Research), project PFS-2015-01. The work of D.T. and L.H. is funded by the Belgian Federal Science Policy through the Interuniversity Attraction Pole P7/37.

\end{document}